# OPTIMIZATION OF A TWO-HOP NETWORK WITH ENERGY CONFERENCING RELAYS

Sharief Abdel-Razeq[1], Ming Zhao[2], Shengli Zhou[1], Zhengdao Wang[3]

[1] Dept. of Electrical and Computer Engineering, University of Connecticut, Storrs, CT, USA

[2] Dept. of Electronic Engineering and Inf. Sci., Univ. of Sci. and Tech. of China, Hefei, China

[3] Dept. of Electrical and Computer Engineering, Iowa State University, Ames, IA, USA

## ABSTRACT

*This paper considers a two-hop network consisting of a source, two parallel half-duplex relay nodes, and two destinations. While the destinations have an adequate power supply, the source and relay nodes rely on harvested energy for data transmission. Different from all existing works, the two relay nodes can also transfer their harvested energy to each other. For such a system, an optimization problem is formulated with the objective of maximizing the total data rate and conserving the source and relays transmission energy, where any extra energy saved in the current transmission cycle can be used in the next cycle. It turns out that the optimal solutions for this problem can be either found in a closed-form or through one-dimensional searches, depending on the scenario. Simulation results based on both the average data rate and the outage probability show that energy cooperation between the two relays consistently improves the system performance.*

## KEYWORDS

## 1. INTRODUCTION

In the past few years, there has been a significant research progress on energy harvesting (EH) communications as it's a promising approach to realize green communications, which allows to power the communication devices and networks with renewable energy sources; see recent review papers [1]–[4] and references therein. Various types of energy sources can be utilized to supplement energy supplies such as solar, wind, vibration, motion, and electromagnetic (EM) wave [5]. Further, through power transfer by radio waves, energy cooperation allows wireless nodes to intentionally transfer some energy to others to assist communications [6].

In EH, the main focus is on the development of energy harvesting models, protocols, and transmission schemes. For instance, in point-to-point communications, both the transmitter and the receiver could be equipped with energy-harvesting devices, and energy transfer can happen between the transceivers [7]–[10].

Recently, considerable research efforts have been extended toward energy harvesting networking like cooperative networks, cognitive radio networks, multiuser interference networks, and cellular networks [11]–[19]. For instance, in [12] a two-hop relay channel with energy-harvesting source and relay nodes, and one-way energy transfer from the source node to the relay node was studied. Multiple access and two-way channels are considered in [13] with energy harvesting transmitters that transfer energy to each other. An energy-harvesting diamond relay channel is analyzed in [14], where the source can transfer some of its harvested energy to the relays. In [18], a multi-relay network is investigated where the relays that are randomly located within the cooperating area and a single best relay is selected to the decoded source message to the destination. Recently, EH in the domain of the fifth generation (5G) has been studied in [19].

There are several other works in the aforementioned research areas related to energy cooperation (EC) but not on energy-harvesting sensor networks [20]–[25]. The transmitter design for wireless information and EC in a multiple-input single-output interference channel was investigated in [20]. In the context of cognitive radio networks, the primary transmitter can transmit power to secondary transmitters such that the latter can obtain the extra power to help the former besides serving their own secondary users [21]. In a wireless powered cellular communication network, downlink wireless energy transfer can be used to assist the uplink information transmissions [22], and EC among base stations has been studied in [24].

Several objectives have been considered when designing energy harvesting communications systems, including data rate, outage probability, harvested energy, and total power consumption [26]. For example, the minimization of the outage probability is considered in [27]–[29] while the data rate is maximized in [30], [31]. The work in [30] attempts to maximize the data rate under heavy channel fluctuations and energy variations. In [31], the optimal water level for the data rate maximization was proposed based on a recursive water-filling approach.

On the other hand, regarding the EH models and based on the availability of non-causal knowledge about energy arrivals at the transmitters, the researchers primarily divide those models into two streams: deterministic models [32], [33] and stochastic models [34], [35]. In the former one, a full knowledge of energy arrival instants and amounts available at the transmitters beforehand. In the stochastic models, the energy renewal processes are regarded as random processes.

For energy scheduling designs, there are two approaches: *offline* and *online*, depending on whether the knowledge of channel state information (CSI) and energy state information (ESI) are available at the beginning of a transmission. In offline approaches, the full (causal and non-causal) knowledge of CSI and ESI during the energy scheduling period is available at the transmitter side a priori and the optimization problems are formulated to maximize certain short-term objectives and solved by convex optimization techniques [31], [36]. Online approaches, on the other side, only account for the causal knowledge of the CSI and ESI [37], [38].

In this paper, we consider a two-hop network consisting of a source, two parallel half-duplex relay nodes, and two destinations, where the two energy-harvesting relays can exchange their harvested energies to each other. To the best of our knowledge, a system with energy-conferencing relays has not been studied in all existing works, and hence the study herein offers a fresh perspective. For such a system, we formulate an optimization problem to maximize the total data rate of the network while conserving the system resources via judicious choices of the source and relays transmission energy on the source-to-relay links and relay-to-destination links, respectively, and the energy transfer between the two relay nodes. The optimal system solution is found either in a closed-form or through one-dimensional searches. Moreover, we study the outage probability at the two destinations and show how EC between the two relays can reduce the system outage.

The remainder of this paper is organized as follows. Section 2 describes the system model and assumptions and Section 3 provides the problem formulation. The solutions to the optimization problem under different scenarios are detailed in Section 4, and a numerical example is presented in Section 5. Section 6 extends the analysis to the outage probability while Section 7 concludes this paper.

## 2. SYSTEM DESCRIPTION AND ASSUMPTIONS

Fig. 1 shows the considered system, which consists of a source S, two parallel relays $R_1$ and $R_2$, and two destinations $D_1$ and $D_2$. Each node is equipped with a single antenna. S, $R_1$, and $R_2$ rely on harvested energy while $D_1$ and $D_2$ are powered with adequate power supply. There is no direct link between S and $D_1$ and $D_2$, and that $R_1$ and $R_2$ are half-duplex working in either a receive mode or in a transmit mode. $R_1$ and $R_2$ will apply decode-and-forward (DF) relaying scheme to forward the data just received from the previous time slot.

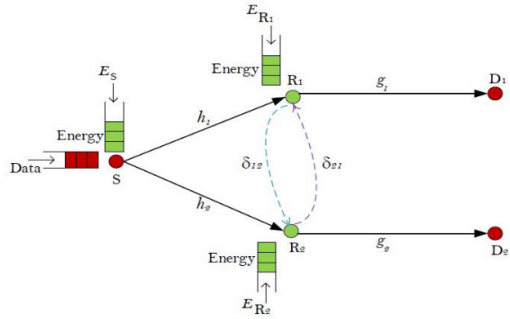

Fig. 1. A two-hop network with energy conferencing relays.

One transmission cycle consists of two stages; the information collection stage, in which, S, $R_1$, and $R_2$ collect the information about channels states and energy harvested at each node then decide the optimum solutions to be used. Once this stage is over, data transmission stage starts which will be done in two consecutive time slots as follows: In the first time slot, S will transmit two different data streams to $R_1$ and $R_2$ simultaneously. In the second time slot, $R_1$ and $R_2$ will decode and then forward their data separately to their destinations. Moreover, data transmission through the upper hop links, i.e., S − $R_1$ and $R_1$ − $D_1$ will be orthogonal on data transmission through the lower hop links, S − $R_2$ and $R_2$ − $D_2$. After finishing a whole transmission cycle, there is a sleeping period, during which, S, $R_1$, and $R_2$ will have the time to harvest more energy to be used in next transmission cycle while $D_1$ and $D_2$ will be idle as shown in Fig. 2.

Throughout the paper, the following set of assumptions are considered.

1) EC between $R_1$ and $R_2$ is done via two conferencing links that could be wired or wireless. These links assumed to be orthogonal to each other and also orthogonal to the source-to-relays and relays-to-destinations links. Moreover, data transmission and energy transfer channels are orthogonal, i.e., energy transfer does not create interference to data communication [39].

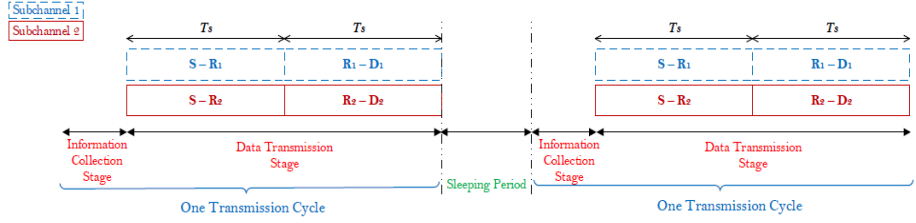

Fig. 2. A timeline showing two consecutive transmission cycles, separated by a sleeping period.

2) S, $R_1$, and $R_2$ have separate batteries and harvested energies are stored in the batteries. The energy loss during the energy transfer procedure will be modelled by a multiplicative efficiency factor, although other models would be applicable.

3) The processing energy required by circuits at any relay is negligible compared to the energy used for signal transmission especially when transmission distances are large which is applicable in our work [40], [41].

4) In our work, the CSI and the energy harvested at S, $R_1$, and $R_2$ nodes are assumed to be collected and known before the start of the transmission cycle and this knowledge assumed to be correct. Hence system optimization can be performed.

## 3. PROBLEM FORMULATION

The delivery of the data to $D_1$ and $D_2$ from S will be done in two time slots; during the first time slot, S will transmit data to $R_1$ and $R_2$ concurrently on two separate channels. In the second time slot, $R_1$ and $R_2$ will forward the received the data to their designated destinations on two separate channels. Note that the $S - R_1$ and $R_1 - D_1$ could be on the same channel, and so does $S - R_2$ and $R_2 - D_2$. On the first hop, S broadcasts independent data streams to $R_1$ and $R_2$. The baseband discrete-time channel with two relays is

$$y_k[m] = h_k x_k[m] + w_k[m], \quad k = 1, 2, \tag{1}$$

where $x_k[m]$ is the signal intended for relay $k$ at time slot $m$, $h_k$ denote the channel gain from the source to relay $k$, $y_k[m]$ is the received signal at relay $k$ during time slot $m$, and $w_k[m]$ is the noise at relay $k$ during time slot m which is assumed independent and identically distributed (i.i.d.) complex Gaussian with $\mathcal{N}(0, \sigma_{w_k}^2)$.

On the second hop, $R_1$ and $R_2$ are responsible to decode the data received from the first hop and then forward it during the next time slot. Hence

$$z_{D_l}[n] = g_l x_l[n] + \bar{w}_l[n], \quad l = 1, 2, \tag{2}$$

where $x_l[n]$ is the signal intended from relay $k$ to destination $D_l$ at time $n$, $g_l$ denote the channel gain from the relay $k$ to destination $D_l$, $z_{D_l}[n]$ is the received signal at destination $D_l$ during time slot $n$, and $\bar{w}_l[n]$ is the noise at destination $D_l$ assumed i.i.d. complex Gaussian with $\mathcal{N}(0, \sigma_{\bar{w}_k}^2)$.

Let $E_S$, $E_{R_1}$, and $E_{R_2}$ denote the energy available per symbol during the data transmission slot at the beginning of each transmission cycle at S, $R_1$, and $R_2$, respectively. Note that $E_S$, $E_{R_1}$ and $E_{R_2}$ are random variables, but the values of their realizations are assumed known before the cycle starts. When $R_1$ transfers $\delta_{12}$ amount of energy to $R_2$, the energy at the receiver is $\gamma_{12}\,\delta_{12}$, where $\gamma_{12}$ is the transfer efficiency from $R_1$ to $R_2$. When $R_2$ transfers $\delta_{21}$ amount of energy to $R_1$, the received energy is $\gamma_{21}\,\delta_{21}$, where $\gamma_{21}$ is the transfer efficiency from $R_2$ to $R_1$. The transfer efficiencies are less than one, which accounts for the potential loss due to various reasons in the energy transfer procedure. Furthermore, $\gamma_{12}$ and $\gamma_{21}$ are not necessarily the same.

On the first hop, define $C_{S,R_1}$ and $C_{S,R_2}$ as the maximum data rates from S to $R_1$ and from S to $R_2$, respectively. Hence

$$C_{S,R_1} = \log_2\left(1 + \frac{|h_1|^2 e_{s_1}}{\sigma_{w_1}^2}\right),\tag{3}$$

$$C_{S,R_2} = \log_2\left(1 + \frac{|h_2|^2 e_{s_2}}{\sigma_{w_2}^2}\right),\tag{4}$$

where $e_{s_1}$ and $e_{s_2}$ is the average energy per transmit symbol from S to $R_1$ and $R_2$, respectively, with $0 \leq e_{s_1} + e_{s_2} \leq E_S$.

On the second hop, define $C_{R_1,D_1}$ and $C_{R_2,D_2}$ as the maximum data rate from $R_1$ to $D_1$ and from $R_2$ to $D_2$, respectively. Hence,

$$C_{R_1,D_1} = \log_2\left(1 + \frac{|g_1|^2 e_{R_1}}{\sigma_{w_1}^2}\right),\tag{5}$$

$$C_{R_2,D_2} = \log_2\left(1 + \frac{|g_2|^2 e_{R_2}}{\sigma_{w_2}^2}\right),\tag{6}$$

where $e_{R_1}$ and $e_{R_2}$ is the average energy per transmit symbol from $R_1$ and $R_2$, respectively.

If $C_1$ and $C_2$ are the maximum data rates of the upper and lower hops' links, respectively. Then, the total data rate of a two-hop DF network is

$$C_{total} = C_1 + C_2 = \min\{C_{S,R_1}, C_{R_1,D_1}\} + \min\{C_{S,R_2}, C_{R_2,D_2}\}.\tag{7}$$

For each direction of energy transfer, one optimization problem needs to be formulated as follows [42].

### 1) From $R_1$ to $R_2$ :

$$\max_{e_{s_1}, e_{s_2}, \delta_{12}, e_{R_1}, e_{R_2}} \quad \mathcal{C}_{\text{total}} \tag{8a}$$

$$\text{s.t} \quad \mathcal{C}_{R_1, D_1} = \mathcal{C}_{S, R_1}, \tag{8b}$$

$$\mathcal{C}_{R_2, D_2} = \mathcal{C}_{S, R_2}, \tag{8c}$$

$$0 \leq e_{R_1} \leq E_{R_1} - \delta_{12}, \tag{8d}$$

$$0 \leq e_{R_2} \leq E_{R_2} + \gamma_{12}\delta_{12}, \tag{8e}$$

$$0 \leq e_{s_1} + e_{s_2} \leq E_S, \tag{8f}$$

$$0 \leq e_{s_1}, 0 \leq e_{s_2}, \tag{8g}$$

$$0 \leq \delta_{12} \leq E_{R_1}. \tag{8h}$$

### 2) From $R_2$ to $R_1$ :

$$\max_{e_{s_1}, e_{s_2}, \delta_{21}, e_{R_1}, e_{R_2}} \quad \mathcal{C}_{\text{total}} \tag{9a}$$

$$\text{s.t} \quad \mathcal{C}_{R_1, D_1} = \mathcal{C}_{S, R_1}, \tag{9b}$$

$$\mathcal{C}_{R_2, D_2} = \mathcal{C}_{S, R_2}, \tag{9c}$$

$$0 \leq e_{R_1} \leq E_{R_1} + \gamma_{21}\delta_{21}, \tag{9d}$$

$$0 \leq e_{R_2} \leq E_{R_2} - \delta_{21}, \tag{9e}$$

$$0 \leq e_{s_1} + e_{s_2} \leq E_S, \tag{9f}$$

$$0 \leq e_{s_1}, 0 \leq e_{s_2}, \tag{9g}$$

$$0 \leq \delta_{21} \leq E_{R_2}. \tag{9h}$$

Note that the equality constraints in (8b) and (8c), and also (9b) and (9c), are imposed since when the maximum total data rate is achieved, any extra energy due to the imbalance of data rates on the first and second hops can be saved for next cycle of transmission and will be added to the newly harvested energy by that node. We will call this strategy as energy saving strategy (ESS). Hence, at the end of any transmission cycle, the saved energy at $R_1$, $R_2$, and S for the next transmission cycle is defined as follows

$$E_{R_i}^{\text{saved}} = E_{R_i} - e_{R_i}^*, \quad i = 1, 2, \tag{10}$$

$$E_S^{\text{saved}} = E_S - (e_{s_1}^* + e_{s_2}^*). \tag{11}$$

where $^*$ indicates the optimal value to be found later. For each realization of the channel gains and the harvested energy levels, the system aims to maximize $\mathcal{C}_{\text{total}}$ while conserving the transmission energy at S, $R_1$, and $R_2$.

The optimization problems in (8a) and (9a) will be carried out separately. The final solution will be selected from the two tentative solutions based on the data rate comparison.

## 4. SOLUTIONS

The optimization problems (8a) and (9a) have multiple inequality constraints. They could be solved through the Karush-Kuhn-Tucker (K.K.T.) conditions, however, the procedure is cumbersome. Through heuristic reasoning, most constraints can be removed in different scenarios and optimal solutions can be found through one-dimensional searches.

### A. First hop is the bottleneck

Assuming that the first hop is the bottleneck and the second hop can fully support the first hop. The sum rate of the first-hop links $C_1^{\text{sum}} = C_{\text{S,R1}} + C_{\text{S,R2}}$ and the problem is to find the energy allotment that maximizes this sum rate subject to the constraint that $e_{s_1} + e_{s_2} = E_S$. This is a standard water-filling problem on power allocation over parallel Gaussian channels [43]. The solution is:

$$e_{s_i} = \left( \nu - \frac{\sigma_{w_i}^2}{|h_i|^2} \right)^+, \quad i = 1, 2, \tag{12}$$

where $\nu$ is chosen so that

$$\sum \left( \nu - \frac{\sigma_{w_i}^2}{|h_i|^2} \right)^+ = E_S, \tag{13}$$

where $(x)^+$ denotes the positive part of $x$.

Once we optimized the first hop, we can proceed to check whether the second hop can support the optimal solution from the first hop. We have the following cases of interest:

*Case A1):* $\frac{|h_1|^2 \sigma_{\hat{w}_1}^2 e_{s_1}^*}{|g_1|^2 \sigma_{w_1}^2} < E_{R_1}$ and $\frac{|h_2|^2 \sigma_{\hat{w}_2}^2 e_{s_2}^*}{|g_2|^2 \sigma_{w_2}^2} < E_{R_2}$. Under this condition, there is no need for EC between $R_1$ and $R_2$ as both have enough energy. The optimal solution is:

$$e_{R_1}^* = \frac{|h_1|^2 \sigma_{\hat{w}_1}^2 e_{s_1}^*}{|g_1|^2 \sigma_{w_1}^2}, \quad e_{R_2}^* = \frac{|h_2|^2 \sigma_{\hat{w}_2}^2 e_{s_2}^*}{|g_2|^2 \sigma_{w_2}^2}, \quad \delta_{21}^* = \delta_{12}^* = 0. \tag{14}$$

*Case A2):* $\frac{|h_1|^2 \sigma_{\hat{w}_1}^2 e_{s_1}^*}{|g_1|^2 \sigma_{w_1}^2} < E_{R_1}$ but $E_{R_2} < \frac{|h_2|^2 \sigma_{\hat{w}_2}^2 e_{s_2}^*}{|g_2|^2 \sigma_{w_2}^2} < E_{R_2} + \gamma_{12} \left( E_{R_1} - \frac{|h_1|^2 \sigma_{\hat{w}_1}^2 e_{s_1}^*}{|g_1|^2 \sigma_{w_1}^2} \right)$. Under this condition, $R_1$ will transfer energy to $R_2$ to support the data rate as required by the first hop.

$$e_{R_1}^* = \frac{|h_1|^2 \sigma_{\hat{w}_1}^2 e_{s_1}^*}{|g_1|^2 \sigma_{w_1}^2}, \quad e_{R_2}^* = \frac{|h_2|^2 \sigma_{\hat{w}_2}^2 e_{s_2}^*}{|g_2|^2 \sigma_{w_2}^2}, \quad \delta_{12}^* = \frac{e_{R_2}^* - E_{R_2}}{\gamma_{12}}. \tag{15}$$

*Case A3):* $\frac{|h_2|^2 \sigma_{\hat{w}_2}^2 e_{s_2}^*}{|g_2|^2 \sigma_{w_2}^2} < E_{R_2}$ but $E_{R_1} < \frac{|h_1|^2 \sigma_{\hat{w}_1}^2 e_{s_1}^*}{|g_1|^2 \sigma_{w_1}^2} < E_{R_1} + \gamma_{21} \left( E_{R_2} - \frac{|h_2|^2 \sigma_{\hat{w}_2}^2 e_{s_2}^*}{|g_2|^2 \sigma_{w_2}^2} \right)$. Under this condition, $R_2$ will transfer energy to $R_1$, which can support the data rate as required by the first hop.

$$e_{R_1}^* = \frac{|h_1|^2 \sigma_{\hat{w}_1}^2 e_{s_1}^*}{|g_1|^2 \sigma_{w_1}^2}, \quad e_{R_2}^* = \frac{|h_2|^2 \sigma_{\hat{w}_2}^2 e_{s_2}^*}{|g_2|^2 \sigma_{w_2}^2}, \quad \delta_{21}^* = \frac{e_{R_1}^* - E_{R_1}}{\gamma_{21}}. \tag{16}$$

**B. First hop is not the bottleneck**

Since the first hop is not the bottleneck, the second hop needs to exhaust all the available energy, i.e., $E_{R_1}^{saved} = E_{R_2}^{saved} = 0$. Moreover, as we have two directions of energy transfer, let us investigate each direction separately.

**1) Energy Transfer from $R_1$ to $R_2$:** Under the assumption that the first hop has enough energy to support the second hop, the target then is to maximize $C_2^{sum} = C_{R_1,D_1} + C_{R_2,D_2}$ on the second hop. Since

$$C_2^{sum} = \log_2\left(1 + \frac{|g_1|^2(E_{R_1} - \delta_{12})}{\sigma_{w_1}^2}\right)$$
$$+ \log_2\left(1 + \frac{|g_2|^2(E_{R_2} + \gamma_{12}\delta_{12})}{\sigma_{w_2}^2}\right), \tag{17}$$

by taking the derivative relative to $\delta_{12}$, we obtain

$$\delta_{12}^{unc} = \frac{\gamma_{12}|g_2|^2\sigma_{w_1}^2 - |g_1|^2\sigma_{w_2}^2}{2\gamma_{12}|g_1|^2|g_2|^2} + \frac{\gamma_{12}E_{R_1} - E_{R_2}}{2\gamma_{12}}. \tag{18}$$

This objective function is convex, which can be verified by checking its second derivative. However, the constraints in (8h) should be imposed to make sure that the solution is in range, we define

$$\bar{\delta}_{12}^{unc} = \begin{cases} 0, & \delta_{12}^{unc} < 0 \\ \delta_{12}^{unc}, & 0 \leq \delta_{12}^{unc} \leq E_{R_1} \\ E_{R_1}, & \delta_{12}^{unc} > E_{R_1}. \end{cases} \tag{19}$$

Now we need to check if condition (8f) has been satisfied by using the energy transfer $\bar{\delta}_{12}^{unc}$. Plugging $\delta_{12} = \bar{\delta}_{12}^{unc}$ into (15) leads to one unique optimal values for $e_{s_1}$ and $e_{s_1}$, which are denoted as $\bar{e}_{s_1}^{unc}$ and $\bar{e}_{s_2}^{unc}$.

If indeed $\bar{e}_{s_1}^{unc} + \bar{e}_{s_2}^{unc} \leq E_S$, then the condition is satisfied which means that the first hop can fully support the optimized second hop, and hence $e_{s_1}^* = \bar{e}_{s_1}^{unc}$ and $e_{s_2}^* = \bar{e}_{s_2}^{unc}$. In short, if *Cases A1-A3* are not applicable, this leads us to where the optimization problem in (8a) is solved as follows.

*Case B1):* If $\bar{e}_{s_1}^{unc} + \bar{e}_{s_2}^{unc} \leq E_s$, then the optimal solution to (8a) is

$$e_{s_1}^* = \bar{e}_{s_1}^{unc}, \ e_{s_2}^* = \bar{e}_{s_2}^{unc}, \ \delta_{12}^* = \bar{\delta}_{12}^{unc}. \tag{20}$$

and

$$e_{R_1} = E_{R_1} - \delta_{12}, \ e_{R_2} = E_{R_2} + \gamma_{12}\delta_{12}. \tag{21}$$

However, if $e_{s_1} + e_{s_2} > E_S$ then the condition (8f) needs to be enforced and a one-dimensional search on $\delta_{12}^*$ solves the optimization problem in (8a) which leads to maximize $C_{total}$ for this direction.

*Case B2):* The solution is

$$e_{R_1} = E_{R_1} - \delta_{12}, \ e_{R_2} = E_{R_2} + \gamma_{12}\delta_{12}, \ \delta_{12}^* = \delta_{12}^{search}. \tag{22}$$

**2) Energy Transfer from R₂ to R₁:** The optimization problem in (9a) is solved following the same steps as in the opposite direction, i.e., $R_1$ to $R_2$. Explanations are now skipped, with only key equations provided. The optimum value of $\delta_{21}$ for an unconstrained optimization on the second hop is

$$\delta_{21}^{\text{unc}} = -\frac{|g_2|^2\sigma_{\text{w}_1}^2 - \gamma_{21}|g_1|^2\sigma_{\text{w}_2}^2}{2\gamma_{21}|g_1|^2|g_2|^2} - \frac{E_{\text{R}_1} - \gamma_{21}E_{\text{R}_2}}{2\gamma_{21}}, \tag{23}$$

based on which we define the truncated version $\bar{\delta}_{21}^{\text{unc}}$ within the interval $[0, E_{\text{R}_2}]$. Plugging $\delta_{21} = \bar{\delta}_{21}^{\text{unc}}$ into (16), and denote the solution on $e_{\text{s}_1}$ and $e_{\text{s}_2}$ as $\bar{e}_{\text{s}_1}^{\text{unc}}$ and $\bar{e}_{\text{s}_2}^{\text{unc}}$. Plugging $e_{\text{s}_1} + e_{\text{s}_2} = E_{\text{S}}$ into (9f) to make sure that this condition is satisfied.

*Case B3):* The optimal solution to (9a) is

$$e_{\text{s}_1}^* = e_{\text{s}_1}^{\text{unc}}, \quad e_{\text{s}_2}^* = e_{\text{s}_2}^{\text{unc}}, \quad \delta_{21}^* = \bar{\delta}_{21}^{\text{unc}} \tag{24}$$

and

$$e_{\text{R}_1} = E_{\text{R}_1} + \gamma_{21}\delta_{21}, \quad e_{\text{R}_2} = E_{\text{R}_2} - \delta_{21}. \tag{25}$$

If $e_{\text{s}_1} + e_{\text{s}_2} > E_{\text{S}}$, a one-dimensional search on $\delta_{21}^*$ will be performed to solve the optimization problem in (9a) which maximizes $C$total for this direction.

*Case B4):* The solution is

$$e_{\text{R}_1} = E_{\text{R}_1} + \gamma_{21}\delta_{21}, \quad e_{\text{R}_2} = E_{\text{R}_2} - \delta_{21}, \quad \delta_{21}^* = \delta_{21}^{\text{search}}. \tag{26}$$

Based on the results from (22) and (26), the system should be able to determine the direction of energy transfer that maximizes the total data rate of the network.

## 5. NUMERICAL EXAMPLE

For numerical simulation, both first- and second-hop channels experience Rayleigh fading as: $h_i \sim \mathcal{CN}\left(0, \sigma_{\text{h}_i}^2\right)$ and $g_i \sim \mathcal{CN}\left(0, \sigma_{\text{g}_i}^2\right)$, $i = 1, 2$. For the first hop, we assume $\sigma_{\text{h}_1}^2 = \sigma_{\text{h}_2}^2$, $\sigma_{\text{w}_1}^2 = \sigma_{\text{w}_2}^2$, and the maximum average SNR at the relay is $\sigma_{\text{h}_1}^2 E_{\text{S}} / \sigma_{\text{w}_1}^2$.

The energy levels $E_{\text{S}}, E_{\text{R}_1}$ and $E_{\text{R}_2}$ at the S, $R_1$, and $R_2$ are randomly generated from a Gaussian distribution $E_{\text{S}} \sim \mathcal{N}\left(\mu_{\text{S}}, \sigma_{E_{\text{S}}}^2\right)$ and $E_{\text{R}_i} \sim \mathcal{N}\left(\mu_{\text{R}_i}, \sigma_{E_{\text{R}_i}}^2\right)$, $i = 1, 2$, respectively. For the second hop, we also assume $\sigma_{\text{g}_1}^2 = \sigma_{\text{g}_2}^2$, and define the average SNR at the destinations as follows

$$\overline{\text{SNR}} = \frac{\overline{\text{SNR}}_{\text{D}_1} + \overline{\text{SNR}}_{\text{D}_2}}{2} = \frac{1}{2}\left[\sigma_{\text{g}_1}^2 \frac{\mu_{\text{R}_1}}{\sigma_{\text{w}_1}^2} + \sigma_{\text{g}_2}^2 \frac{\mu_{\text{R}_2}}{\sigma_{\text{w}_2}^2}\right]. \tag{27}$$

The numerical values of the system variables are set as follows: $\sigma_{\text{w}_1}^2 = \sigma_{\text{w}_2}^2 = 1\text{mJ}$, $\sigma_{\text{w}_1}^2 = \sigma_{\text{w}_2}^2 = 1\text{mJ}$, and $\gamma_{12} = \gamma_{21} = 90\%$ under the assumption that the conferencing links are wires. It is worth to mention here that if these links are assumed to be wireless, then $\gamma_{12}$ and $\gamma_{21}$ should be much lower due to high energy loss in the free space.

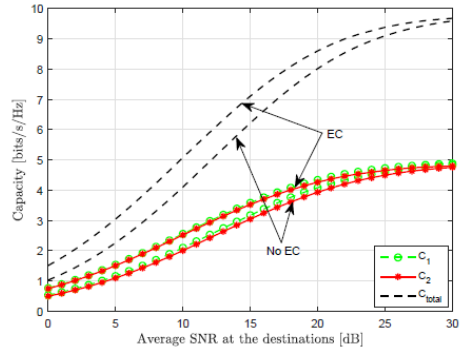

Fig. 3. Capacity versus the average SNR at the destinations for energy cooperation (EC) and no energy cooperation (No EC), where the relay energy levels and the channel gains are randomly generated with $E_{R_1}, E_{R_2}, E_S \sim \mathcal{N}(100\text{mJ}, 50\text{mJ})$.

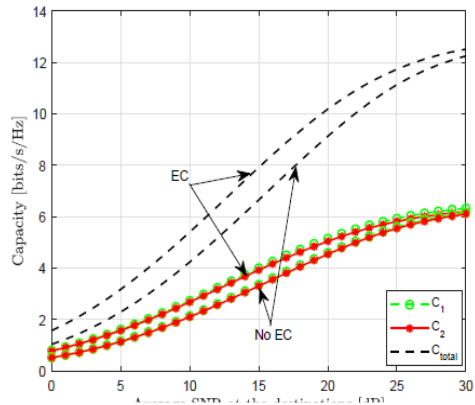

Fig. 4. Capacity versus the average SNR at the destinations for EC and No EC, where the relay energy levels and the channel gains are randomly generated with $E_S \sim \mathcal{N}(300\text{mJ}, 50\text{mJ})$ while $E_{R_1}, E_{R_2} \sim \mathcal{N}(100\text{mJ}, 50\text{mJ})$.

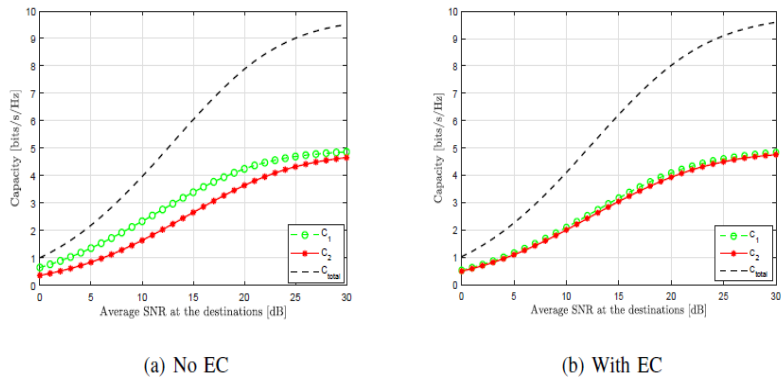

(a) No EC        (b) With EC

Fig. 5. Capacity versus the average SNR at the destinations, where the relay energy levels and the channel gains are randomly generated with $E_{R_1} \sim \mathcal{N}(200\text{mJ}, 50\text{mJ})$ while $E_{R_2}, E_S \sim \mathcal{N}(100\text{mJ}, 50\text{mJ})$.

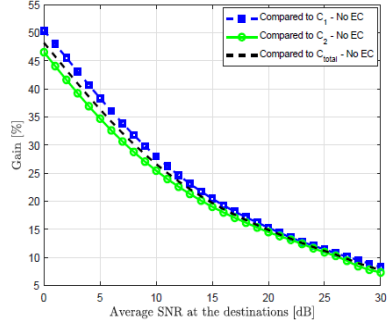

Fig. 6. Gain versus the average SNR at the destinations, where the relay energy levels and the channel gains are randomly generated with $E_{R_1}, E_{R_2}, E_S \sim \mathcal{N}(100mJ, 50mJ)$.

Fig. 3, Fig. 4, and Fig. 5 show capacity as a function of the average SNR at the destinations with ESS is adopted for different values of $E_{R_1}, E_{R_2}, E_S$. Fig. 3 and Fig. 4 show that, for any values of $E_{R_1}, E_{R_2}, E_S$, EC is always helpful for the system and better than no EC. It is worth to mention here that, the role of ES is to control the maximum capacity that system can reach, which is expected. In Fig. 3, the highest capacity was about 10 bits/s/Hz while this floor got higher to about 12 bits/s/Hz when ES is increased as Fig. 4 shows.

On the other hand, Fig. 5 shows that, when $E_{R_2}$ is less than $E_{R_1}$, EC will be helpful to increase the data rate at $D_2$ with no degradation on $D_1$'s data rate. In other words, the strong relay will not sacrifice its rate by EC but it will help the weak relay which benefits the overall system.

Fig. 6 shows the gain we get by adopting EC. It can be easily observed that this gain is much higher at low average SNR. This is justifiable as, at low average SNR, EC is crucial to help the system to overcome the bad channel states. However, this gain will decrease as average SNR increases due to fact that the channel state is getting better and EC is not critically needed any more.

Fig. 7 shows scenarios when a one-dimensional search is needed which corresponds to *Case B3*. In Fig. 7(a) and Fig. 7(b), optimum $\delta_{12}$ happens to be outside the designated range which means that the system should search for a viable solution. The search will start from 'start searching' point and goes backward until it finds $\delta_{12}$ that is in the range and can be supported by the first hop then stops searching. The dashed lines show the searched domain.

In Fig. 7(c), even though the peak occurs within the designated range, the system must perform the search as optimum $\delta_{12}$ couldn't be supported by the first hop due to low value of $E_S$. The search will start from the peak in two directions and stops at the point that can be supported by the first hop.

Fig. 8 shows the percentage of occurrence of each case we mentioned earlier. This figure confirms that all possible scenarios have been covered. Moreover, as expected, this percentage depends on the energy level at each node. In Fig. 8(a), as $E_S$ is less than $E_{R_1}$ and $E_{R_2}$, this means that first hop will be more likely the bottleneck. This is why in this figure *Cases A1-A3* appears more frequently than other cases. On the other hand, in Fig. 8(b), $E_S$ is larger than $E_{R_1}$ and $E_{R_2}$ which means that the first hop is less likely to be the bottleneck. This justifies why *Cases B1-B4* show up more frequently than other cases.

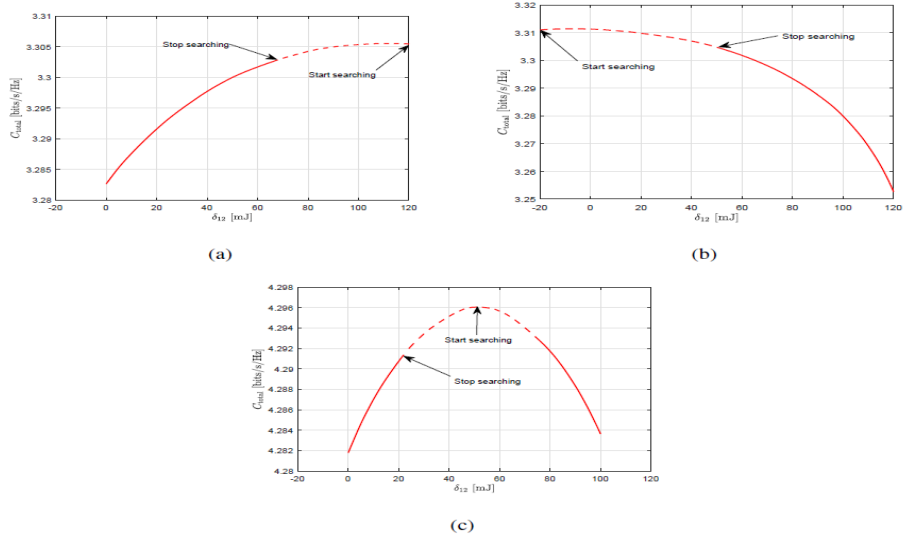

Fig. 7. $C$total versus $\delta_{12}$ for three different realizations drawn from $E_S \sim \mathcal{N}(50\text{mJ}, 20\text{mJ})$ while $E_{R_1}, E_{R_2} \sim \mathcal{N}(100\text{mJ}, 50\text{mJ})$.

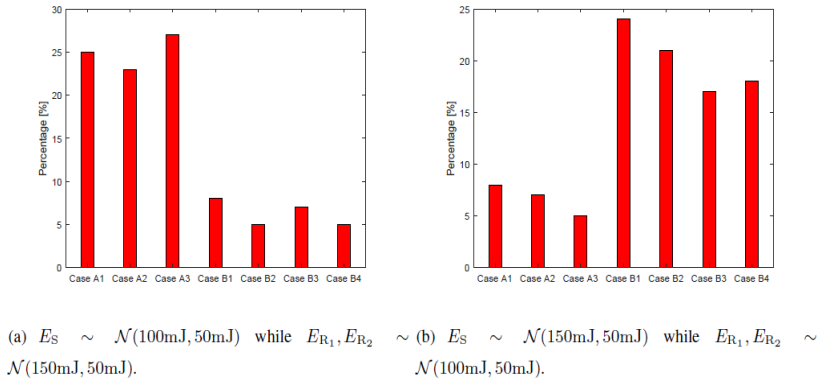

(a) $E_S \sim \mathcal{N}(100\text{mJ}, 50\text{mJ})$ while $E_{R_1}, E_{R_2} \sim \mathcal{N}(150\text{mJ}, 50\text{mJ})$.

(b) $E_S \sim \mathcal{N}(150\text{mJ}, 50\text{mJ})$ while $E_{R_1}, E_{R_2} \sim \mathcal{N}(100\text{mJ}, 50\text{mJ})$.

Fig. 8. Percentage of occurrence of each case when the relay energy levels and the channel gains are randomly generated.

# 6. EXTENSION TO OUTAGE PROBABILITY

In the ideal scenario, S transmits the two data streams to $R_1$ and $R_2$ and they decode the received data perfectly and then forward it to their destinations with no outage. However, this is not the situation in the real scenario, in which, the outage happens at $R_1$, $R_2$, $D_1$ or $D_2$. If $R_1^*$ and $R_2^*$ are the target rates at $D_1$ and $D_2$, respectively, then the optimum system energy will be as follows

$$e_{s_1}^* = \frac{\sigma_{w_1}^2 (2^{R_1^*} - 1)}{|h_1|^2}, \quad e_{s_2}^* = \frac{\sigma_{w_2}^2 (2^{R_2^*} - 1)}{|h_2|^2}, \tag{28}$$

$$e_{R_1}^* = \frac{\sigma_{\tilde{w}_1}^2 (2^{R_1^*} - 1)}{|g_1|^2}, \quad e_{R_2}^* = \frac{\sigma_{\tilde{w}_2}^2 (2^{R_2^*} - 1)}{|g_2|^2}. \tag{29}$$

Now, we can articulate the possible scenarios of interest:

a) If $E_S \geq e_{s_1}^* + e_{s_2}^*$, $e_{R_1}^* \leq E_{R_1}$, and $e_{R_2}^* \leq E_{R_2}$, then no EC needed. However, if $e_{R_1}^* > E_{R_1}$ while $e_{R_2}^* \leq E_{R_2}$, then $\delta_{21}^* = \frac{e_{R_1}^* - E_{R_1}}{\gamma_{21}}$ Joules of energy transfer will happen from $R_2$ to $R_1$ under the condition that $R_2$ can support this transfer. If $e_{R_1}^* \leq E_{R_1}$ and $e_{R_2}^* > E_{R_2}$, then $\delta_{12}^* = \frac{e_{R_2}^* - E_{R_2}}{\gamma_{12}}$ of energy transfer will happen from $R_1$ to $R_2$ under the condition that $R_1$ can support this transfer. Hence, all nodes are working and no outage will happen.

b) If $E_S \geq e_{s_1}^*$ and $e_{R_1}^* \leq E_{R_1}$, then no EC needed from $R_2$. However, if $e_{R_1}^* > E_{R_1}$, then energy transfer will happen from $R_2$ to $R_1$ under the condition that $R_2$ can support $R_1$ with $\delta_{21}^* = \frac{e_{R_1}^* - E_{R_1}}{\gamma_{21}}$ Joules. Hence, $R_2$ and $D_2$ are in outage.

c) If $E_S \geq e_{s_2}^*$ and $e_{R_2}^* \leq E_{R_2}$, then no EC needed from $R_1$. However, if $e_{R_2}^* > E_{R_2}$, then energy transfer will happen from $R_1$ to $R_2$ under the condition that $R_1$ can support $R_2$ with $\delta_{12}^* = \frac{e_{R_2}^* - E_{R_2}}{\gamma_{12}}$ Joules. Hence, $R_1$ and $D_1$ are in outage.

d) If there is no enough energy to support the first and second hop, all nodes will be off during that transmission cycle. Hence, all nodes are in outage.

For the outage probability, we set both target rates $R_1^*$ and $R_2^*$ at 1.5 bits/s/Hz. Fig. 9 shows outage probability as a function of the average SNR at the destinations for different values for EC and No EC scenarios with and without ESS. It can be easily observed that the outage probability will be decreased by adopting EC as the two relays exchanging energy which allows the two destinations to overcome the outage difficulties and to satisfy their requirements. Moreover, this figure shows the benefit of exploiting the idea of saving the extra energy to be used in the next cycle of the transmission cycle.

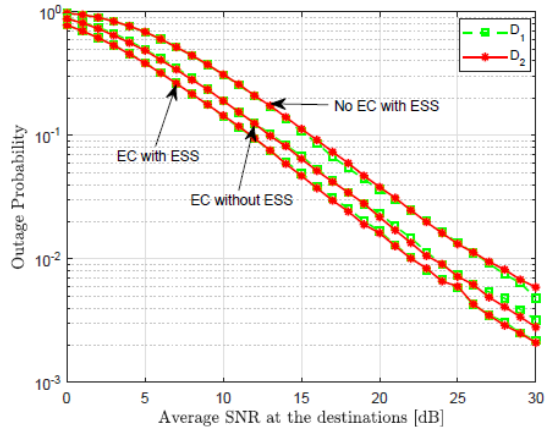

Fig. 9. Outage probability when $E_{R_1}, E_{R_2}, E_S \sim \mathcal{N}(100\text{mJ}, 50\text{mJ})$.

# 7. Conclusions

In this paper, we studied a two-hop network that has two energy harvesting relays which can exchange energy through conferencing links. Via suitable choices of the source and relays transmission energy, and the amount of energy transfer between the two relay nodes, the system data rate is maximized while the system energy is conserved. The actual harvested energy level at the node and the channel state information decide the way how the optimal solutions can be obtained. They can vary from a closed-form solution to one-dimensional searches. Moreover, system outage probability was investigated to show the performance improvement through energy cooperation.